\documentclass[aps,prl,twocolumn,superscriptaddress]{revtex4-2}  
\usepackage{bm}        
\usepackage{amssymb}   

\usepackage{color}
\usepackage{graphicx}
\usepackage{soul}
\usepackage{subfigure,amsmath, amsthm, amssymb}
\usepackage[colorlinks,citecolor=red]{hyperref}
\usepackage{csquotes}

\newcommand{\be}{\begin{equation}}
\newcommand{\ee}{\end{equation}}
\newcommand{\bea}{\begin{eqnarray}}
\newcommand{\eea}{\end{eqnarray}}

\begin{document}

\title{Mpemba effect in an anisotropically driven granular gas}

\author{Apurba Biswas}
\email{apurbab@imsc.res.in}
\affiliation{The Institute of Mathematical Sciences, C.I.T. Campus, Taramani, Chennai 600113, India}
\affiliation{Homi Bhabha National Institute, Training School Complex, Anushakti Nagar, Mumbai 400094, India}
\author{V . V. Prasad}
\email{prasad.vv@cusat.ac.in}
\affiliation{Government Arts and Science College, Nadapuram, Kozhikode 673506, India}
\affiliation{Department of Physics, Cochin University of Science and Technology,  Kochi 682022, India}
\author{R. Rajesh} 
\email{rrajesh@imsc.res.in}
\affiliation{The Institute of Mathematical Sciences, C.I.T. Campus, Taramani, Chennai 600113, India}
\affiliation{Homi Bhabha National Institute, Training School Complex, Anushakti Nagar, Mumbai 400094, India}

\begin{abstract}
We demonstrate the existence, as well as determine the conditions, of a Mpemba effect -  a counterintuitive phenomenon where a hotter system equilibrates faster than a cooler system when quenched to a cold temperature - in anisotropically driven granular gases. In contrast to earlier studies of Mpemba effect in granular systems, the initial states are stationary, making it a suitable system to experimentally study the effect. Our theoretical predictions for the regular, inverse and strong Mpemba effects agree well with results of event-driven molecular dynamics simulations of hard discs.
\end{abstract}

\maketitle
When two identical physical systems, initially at two different temperatures, are quenched to the same lower temperature, one would intuitively expect that the hotter system takes a longer time to equilibrate. However, there exists a counter-intuitive observation that hot water may freeze faster than colder water, as documented centuries ago by Aristotle~\cite{aristotle,1952meteorologica}, and referred to as Mpemba effect after Mpemba and Osborne~\cite{Mpemba_1969} who quantified the phenomena. Although the exact mechanism for such anomalous behaviour found in water is not known, several candidates have been proposed:  evaporation~\cite{Mirabedin-evporation-2017}, convection~\cite{vynnycky-convection:2015}, supercooling~\cite{david-super-cooling-1995}, dissolved gases~\cite{katz2009hot}, hydrogen bonding~\cite{zhang-hydrbond1-2014,tao-hydrogen-2017,Molecular_Dynamics_jin2015mechanisms} and non-equipartition of energy among the different degrees of freedom~\cite{gijon2019paths}. The Mpemba effect has also been demonstrated in other physical systems such as clathrate hydrates~\cite{paper:hydrates}, magnetic alloys~\cite{chaddah2010overtaking}, polylactides~\cite{Polylactide} and more recently in colloidal system~\cite{kumar2020exponentially}.  

Theoretical analysis has focused on model spin systems~\cite{PhysRevLett.124.060602,SpinGlassMpemba}, molecular gases~\cite{moleculargas,gonzalez2020mpemba} and three state Markov systems~\cite{Lu-raz:2017,Klich-2019,klich2018solution}. In the third case, the exact condition for the existence of the Mpemba effect could be derived by tracking the distance between probability distributions during the relaxation process. Moreover, such a framework also suggests the existence of an inverse Mpemba effect~\cite{Lu-raz:2017}, where an initially colder system can heat up faster than an initially hotter system, the strong Mpemba effect~\cite{Klich-2019} where an initial state results in an exponentially faster cooling, and optimal heating protocols~\cite{PhysRevLett.124.060602} for such systems. However, these systems are far from experimental realization.

An area where a strong interplay between experiments and analytical calculations are possible, promoting a deeper understanding on the Mpemba effect, is driven granular systems, where the external driving compensates for the loss of kinetic energy in inelastic collisions. The Mpemba effect has been demonstrated for homogeneous driven smooth as well as as rough gases, provided the initial states are allowed to be non-stationary~\cite{Lasanta-mpemba-1-2017,Torrente-rough-2019,mompo2020memory,PhysRevE.102.012906}. The results from an exact analysis of the driven, inelastic Maxwell gas where the rate of collision is taken to be independent of the relative velocity of the colliding particles, is consistent with the above results for the mono-dispersed gas~\cite{PhysRevE.102.012906}. The Mpemba effect has also been demostrated in sheared inertial suspensions~\cite{PhysRevE.103.032901} of inelastic hard spheres given that one of the system is prepared in a quasi-equilibrium state.  However, the lack of stationarity of the initial conditions makes it difficult for experimental realization because while stationary states are easy to achieve due to them being attractive, non-stationary states require careful preparation of the initial state.

In this paper, we present an analysis of an anisotropically driven, mono-dispersed granular gas in two dimensions, and show the existence of the Mpemba effect starting from initial conditions that are stationary. In addition, we derive the conditions for the inverse and the strong Mpemba effect in this system. The analysis is exact for stationary states that are close to the final stationary state. For generic initial stationary states, we verify our theoretical predictions with detailed event-driven molecular dynamics simulations of hard discs. We propose that this set up of anisotropically driven granular gases is ideal for studying the Mpemba effect, and later for possibly more practical applications.

Consider a two-dimensional granular gas composed of identical, smooth, inelastic hard particles. The velocities of these particles change in time through momentum conserving binary collisions. When two particles $i$ and $j$ with velocities $\boldsymbol{v}_i$ and $\boldsymbol{v}_j$ collide,  the new velocities $\boldsymbol{v}'_i$ and $\boldsymbol{v}'_j$ are given by 
\bea
\begin{split}
\boldsymbol{v}'_i=\boldsymbol{v}_i -\frac{1+r}{2}[(\boldsymbol{v}_i-\boldsymbol{v}_j).\boldsymbol{\hat{e}}]\boldsymbol{\hat{e}},  \\
\boldsymbol{v}'_j=\boldsymbol{v}_j +\frac{1+r}{2}[(\boldsymbol{v}_i-\boldsymbol{v}_j).\boldsymbol{\hat{e}}]\boldsymbol{\hat{e}},
\end{split} \label{eq:collision}
\eea
where $r$ is the co-efficient of restitution and $\boldsymbol{\hat{e}}$ is the unit vector along the line joining the centres of the particles at contact. The particles are anisotropically driven at a constant rate. At long times, the system goes into a stationary state.

The velocity distribution function $f(\boldsymbol{v},t)$, defined as the number density of particles having velocity $\boldsymbol{v}$ at time $t$, obeys the Enskog-Boltzmann equation~\cite{Noije:98}
\begin{align}
\frac{\partial}{\partial t}f(\boldsymbol{{v}},t)=\chi I(f,f) + \Big(\frac{\xi^2_{0x}}{2} \frac{\partial^2}{\partial v^2_x} + \frac{\xi^2_{0y}}{2} \frac{\partial^2}{\partial v^2_y}\Big) f(\boldsymbol{{v}},t), \label{boltzmann eq}
\end{align}
where $\chi$~\cite{brilliantov2010kinetic} is the pair correlation function, $I(f,f)$ is the collision integral which accounts for the rate of collision of two particles being proportional to their relative velocity [see  Eq.~(S2) of Supplemental Material (SM)~\cite{supplemental}], and $\xi^2_{0x}$ and $\xi^2_{0y}$ are the variances or strengths of the white noise along the $x$- and $y$-directions respectively. 
 
Note that in Eq.~(\ref{boltzmann eq}), we have introduced different driving strengths along the two directions, and we will refer to such driving as anisotropic driving. The  Enskog-Boltzmann equation can be used to describe a spatial system but here we consider a spatially homogeneous system such that the spatial degrees of freedom are ignored. Moreover, as usually assumed in kinetic theory for dilute gases, we apply the molecular chaos hypothesis to use product measure for the joint velocity distribution function in the collision integral $I(f,f)$~\cite{supplemental}.

The anisotropic driving implies that, though the system remains homogeneous with $\langle\boldsymbol{v}\rangle=0$, the mean kinetic energies per particle (granular temperature) along the two directions are different. The granular temperatures are defined as
$T_i(t)=(2/n)\int d \boldsymbol{v} \frac{1}{2} m v^2_i  f(\boldsymbol{{v}},t)$, where $n=\int d \boldsymbol{v} f(\boldsymbol{{v}},t)$ is the number density, $m$ is the mass of the particles, and $i=x, y$. We introduce two quantities $T_{tot}$ and  $T_{dif}$ which are the total and the difference of the temperatures in the two directions:
\bea
\begin{split}
T_{tot}(t)=T_x(t) + T_y(t),  \\
T_{dif}(t)=T_x(t) - T_y(t).  \label{eq Tt}
\end{split}
\eea

Within kinetic theory, the velocity distribution of a driven granular gas, $f(\boldsymbol{{v}},t)$, is expanded about the Gaussian in terms of Sonine polynomials~\cite{Noije:98,brilliantov2010kinetic}. For the sake of simplicity, we restrict ourselves to the first term in the expansion, namely a Gaussian. Adding more terms will not change the qualitative results obtained in the paper.
Thus, we write
\bea
f(\boldsymbol{v},t)=\frac{mn}{2\pi \sqrt{T_x(t) T_y(t)}}\exp\left[ -\frac{m v^2_{x}}{2 T_x(t)}-\frac{m v^2_{y}}{2 T_y(t)} \right]. \label{gaussian distribution}
\eea
Substituting Eq.~(\ref{gaussian distribution}) into Eq.~(\ref{boltzmann eq}), we find that the equations governing the temporal evolution of $T_{tot}(t)$ and $T_{dif}(t)$ form a closed set of coupled non-linear differential equations,  and is given by
\bea
\begin{split}
\frac{\partial }{\partial t} T_{tot}(t)=\mathcal{F}(T_{tot},T_{dif}),\\
\frac{\partial }{\partial t}T_{dif}(t)=\mathcal{G}(T_{tot},T_{dif}).
\end{split} \label{time ev}
\eea
The details of the derivation and the functional forms for $\mathcal{F}(T_{tot},T_{dif})$ and $\mathcal{G}(T_{tot},T_{dif})$ are given in SM~\cite{supplemental}. 

We define the Mpemba effect as follows. Consider two systems with different initial $T_{tot}$. Both systems are quenched to the same final stationary state whose temperature is lower. If the system that is initially hotter reaches the final stationary state faster,  then we say that the system shows the Mpemba effect. Likewise, if the initial systems are quenched to a higher temperature and the the cooler system equilibrates faster, we will say that the system shows the inverse Mpemba effect. Finally, for both effects, we will say that the system shows strong Mpemba effect, if the hotter system relaxes exponentially faster than the cooler system for the Mpemba effect and vice versa for the inverse Mpemba effect. We will demonstrate the existence of all these four features for the anisotropically driven granular gas, both analytically using a linearized theory as well as through  
event-driven molecular dynamics simulations.

We first analytically demonstrate the Mpemba effect by considering only those initial stationary states that are close to the final stationary state,  allowing for linearization of the system thus making it tractable. We denote the parameters of the final stationary state by $T^{st}_{tot}$ and $T^{st}_{dif}$. In the stationary state, the time derivatives in Eq.~(\ref{time ev}) can be set to zero and hence $\mathcal{F}(T^{st}_{tot},T^{st}_{dif})=0$ and $\mathcal{G}(T^{st}_{tot},T^{st}_{dif})=0$. Let $\delta T_{tot}(t)=T_{tot}(t)- T^{st}_{tot}$ and $\delta T_{dif}(t)=T_{dif}(t) - T^{st}_{dif}$ denote the time-dependent deviation from the stationary state values. For small deviations, the non-linear differential equations in Eq.~(\ref{time ev}) can be linearized about the stationary state values to give
\begin{equation}
\frac{d}{dt}\begin{bmatrix}
\delta T_{tot}(t) \\
\delta T_{dif}(t)
\end{bmatrix}=-~\boldsymbol{R}\begin{bmatrix}
\delta T_{tot}(t) \\
\delta T_{dif}(t)
\end{bmatrix}, \label{time ev delta T}
\end{equation}
where $\boldsymbol{R}$ is a $2\times2$ matrix with constant entries $R_{ij}$ as given in Eqs.~(S8) of SM~\cite{supplemental}.  $\delta T_{tot}(t)$ and $\delta T_{dif}(t)$ then relaxes in time to zero as
\bea
\begin{split}
\delta T_{tot}(t)=K_+ e^{-\lambda_+ t} + K_- e^{-\lambda_- t}, \\
\delta T_{dif}(t)=L_+ e^{-\lambda_+ t} + L_- e^{-\lambda_- t}, \label{time ev delta Tt}
\end{split}
\eea 
where the coefficients $K_+, K_-, L_+$ and $L_-$ are as given in Eqs.~(S10) of SM~\cite{supplemental},  $\lambda_\pm$ are the eigenvalues of $\boldsymbol{R}$ and $\gamma=\lambda_+-\lambda_-$.

We now derive the condition for the Mpemba effect to be present in the linearized regime,  based on the analysis of Eq.~(\ref{time ev delta Tt}).  Consider two systems $P$ and $Q$ whose stationary state parameters are denoted  as $[T^P_{tot}, T^P_{dif}]$ and $[T^Q_{tot}, T^Q_{dif}]$ respectively. We will assume that  $P$ is hotter than  $Q$, i.e., $T^P_{tot}> T^Q_{tot}$. On quenching to $[T^{st}_{tot}, T^{st}_{dif}]$, the Mpemba effect will be  present when there exists a finite time $\tau$ such $T^P_{tot}(t)<T^Q_{tot}(t)$ for $t >\tau$.  To characterize the difference between the two system, we introduce the quantities $\Delta T_{tot}= T^P_{tot}(0) - T^Q_{tot}(0)$ and $\Delta T_{dif}= T^P_{dif}(0) - T^Q_{dif}(0)$. From Eq.~(\ref{time ev delta Tt}), written for both $P$ and $Q$,  the time $\tau$ at which the two relaxation curves cross, corresponding to $T^P_{tot}(\tau)=T^Q_{tot}(\tau)$, is given by
\bea
\tau=\frac{1}{\gamma} \ln \Big[\frac{R_{12} \Delta T_{dif} - (\lambda_- - R_{11}) \Delta T_{tot}}{R_{12} \Delta T_{dif} - (\lambda_+ - R_{11}) \Delta T_{tot}} \Big]. \label{eq crossing time}
\eea

\begin{figure}
\centering
\includegraphics[width=\columnwidth]{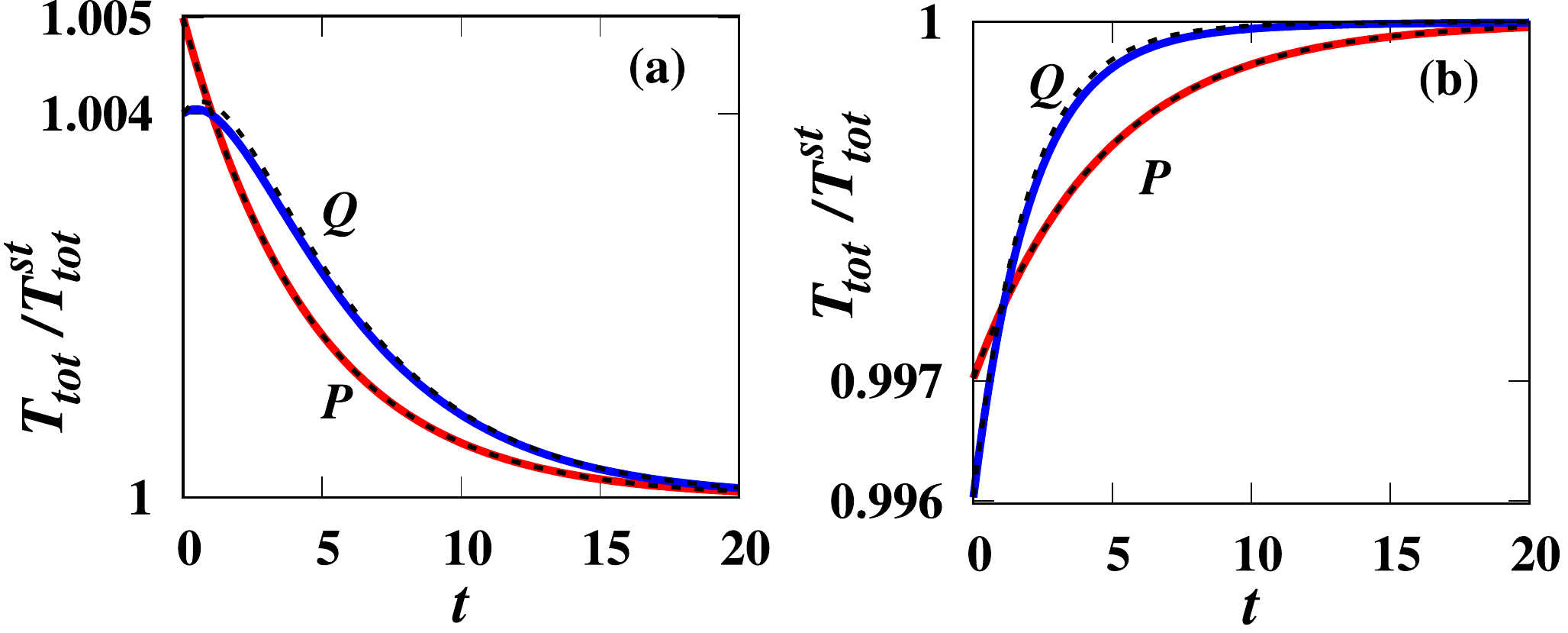}
\caption{(a) The time evolution of $T_{tot}(t)$ with time $t$ for two  systems $P$ and $Q$, with initial conditions $T^P_{tot}(0)$=100.5, $T^P_{dif}(0)$=28, $T^Q_{tot}(0)$=100.4 and $T^Q_{dif}(0)$=$20.01$, show the Mpemba effect when quenched to the final stationary state values of $T^{st}_{tot}$=100 and $T^{st}_{dif}$=28.2.  (b) The initial conditions  $T^P_{tot}(0)$=100.5, $T^P_{dif}(0)$=28, $T^Q_{tot}(0)$=100.4 and $T^Q_{dif}(0)$=$20.01$ show the inverse Mpemba effect when quenched to the final stationary state values of $T^{st}_{tot}$=100.8 and $T^{st}_{dif}$=27.9. The solid lines represent the exact time evolution of $T_{tot}$ and the dashed lines represent its time evolution  after linearization. The other parameters used for the systems are $m$=1, $n$=0.02 and $r$=0.65.  }\label{fig11}
\end{figure}

For the Mpemba effect to be present, we require that $\tau >0$, or equivalently (since $\gamma>0$), the argument of the logarithm in Eq.~(\ref{eq crossing time}) should be greater than one. 
We immediately obtain the criterion for the crossing of the two trajectories as
\begin{equation}
R_{12}\Delta T_{dif}>(\lambda_+-R_{11})\Delta T_{tot}. \label{condition for mpemba}
\end{equation}
In Fig.~\ref{fig11}(a), we choose initial conditions $P$ and $Q$ such that Eq.~(\ref{condition for mpemba}) is satisfied. The trajectories cross at the point as predicted by Eq.~(\ref{eq crossing time}). For initial stationary states that are close to the final state, there is little difference between the  linearized (dotted lines) and the full numerical solutions (solid lines).

\begin{figure}
\centering
\includegraphics[width=\columnwidth]{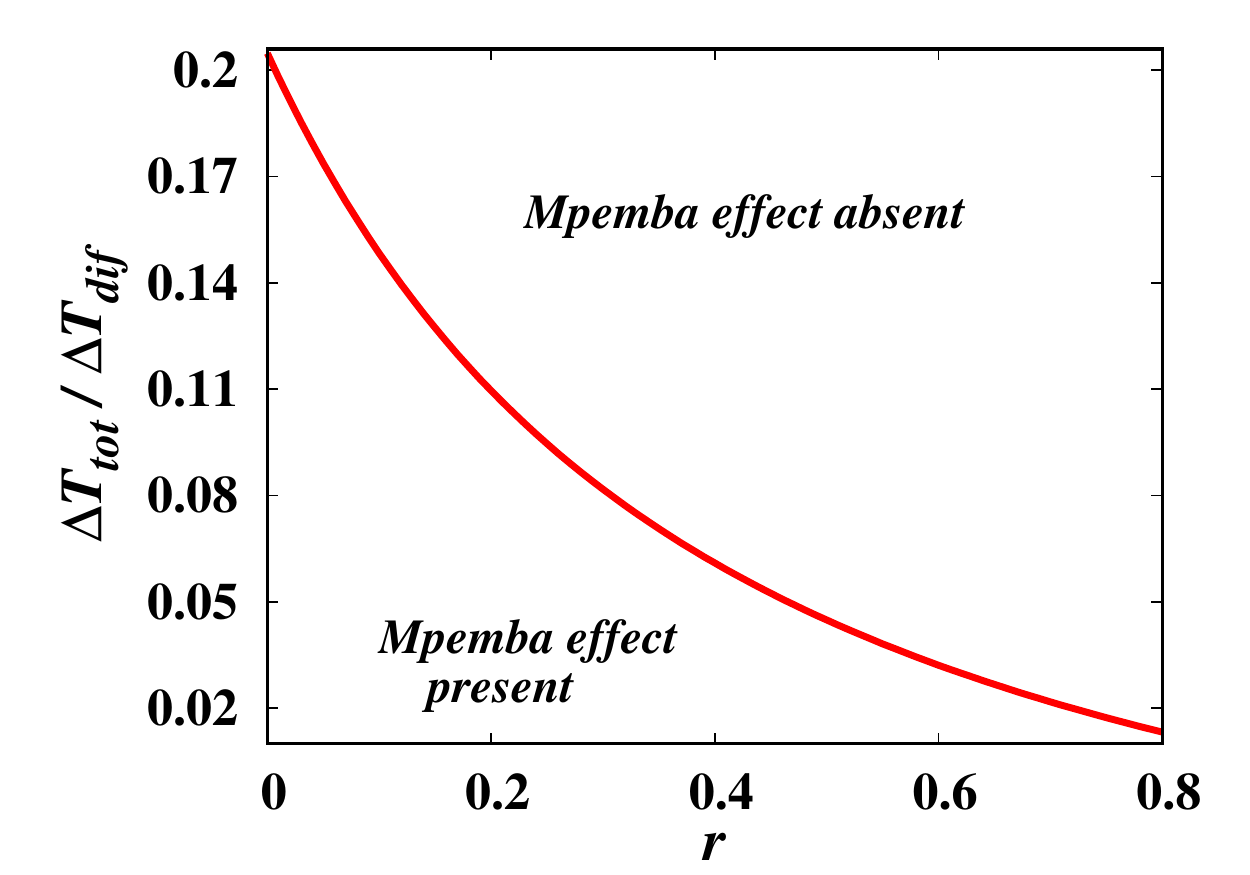}
\caption{The $\Delta T_{tot}/\Delta T_{dif}$--$r$ phase diagram showing regions where the Mpemba effect is observed and $r$ is the coefficient of restitution. All other parameters are kept constant. Here, both the systems are quenched to the final stationary state given by $T^{st}_{tot}=8.0$ and $T^{st}_{dif}=1.2$. The region below the critical line show the Mpemba effect whereas the region on the other side of the critical line does not show the Mpemba effect.}\label{fig12}
\end{figure}

In Fig.~\ref{fig12}, we identify the region of phase space (initial conditions) where the Mpemba effect is observable for varying coefficient of restitution $r$,  based on Eq.~(\ref{condition for mpemba}). The initial conditions in the region below the line in the phase diagram show the Mpemba effect whereas the other region does not show the effect. The phase diagram is for a generic final stationary state with parameters $T^{st}_{tot}=8.0$ and $T^{st}_{dif}=1.2$.  As $r$ approaches unity, the gas become more isotropic, and the the key feature responsible for the presence of the Mpemba effect, i.e., anisotropy of temperatures, is lost and hence the Mpemba effect is not observed.

It can be shown that an inverse Mpemba effect also exists wherein the system is heated instead of being cooled. The condition for the inverse Mpemba effect to be present turns out to be the same as in Eq.~(\ref{condition for mpemba}). An example is illustrated in Fig.~\ref{fig11}(b). Initially $Q$ is at a lower temperature. On being quenched to a common higher temperature, it can be seen that $Q$ equilibrates faster. Again, the difference between the exact linearized solution and the full numerical solution of the non-linear equation is negligible.

\begin{figure}
\centering
\includegraphics[width=\columnwidth]{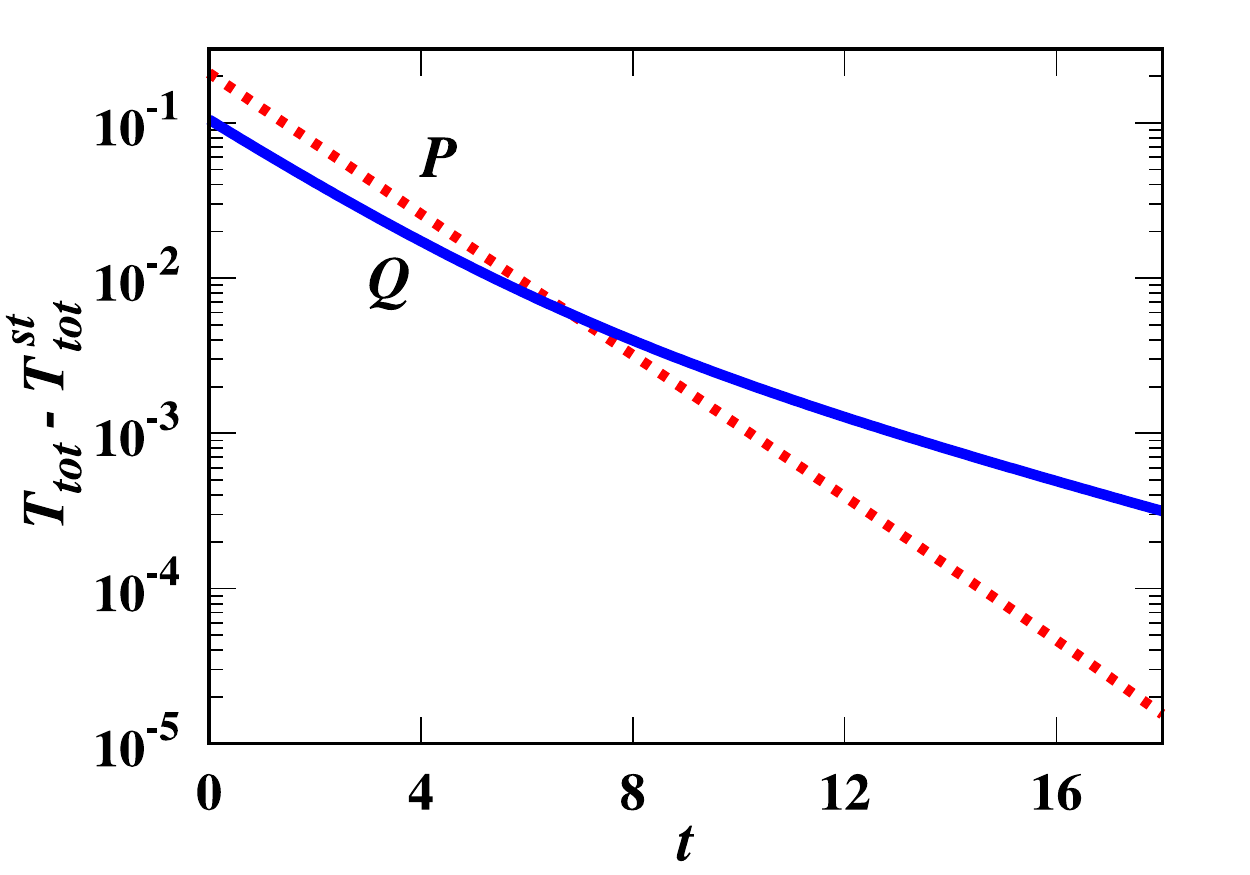}
\caption{Time evolution of $T_{tot}(t)$ with $t$ for two identical systems $P$ and $Q$ with initial conditions $T^P_{tot}(0)$=100.21, $T^P_{dif}(0)$=32.34, $T^Q_{tot}(0)$=100.11 and $T^Q_{dif}(0)$=$30$ which are chosen close to the final stationary state values of $T^{st}_{tot}$=100 and $T^{st}_{dif}$=28.2. $P$ cools exponentially faster than $Q$ though it has higher initial temperature and thus exhibits the strong Mpemba effect.}\label{fig15}
\end{figure}

We also explore the possibility of the strong Mpemba effect in which the system at higher temperature cools exponentially faster. The linear evolution equation in  Eq.~(\ref{time ev delta Tt})  allows certain set of initial conditions to relax to the final stationary state exponentially faster compared to other initial states. The effect may be realized when the coefficient associated with the slower relaxation rate  in the time evolution of total temperature, $T_{tot}(t)$ vanishes. Setting the coefficient associated with the slower relaxation rate  to zero, we obtain the condition for the strong Mpemba effect to be
\begin{align}
T_{tot}(0)=\frac{R_{12}}{\lambda_+ - R_{11}}T_{dif}(0)-c, \label{strong mpemba linear}
\end{align}
where $c=R_{12}/(\lambda_+-R_{11}) T^{st}_{dif} - T^{st}_{tot}$. For a system with all other parameters kept fixed, solution of Eq.~(\ref{strong mpemba linear}) in terms of 
$T_{tot}(0)$ and $T_{dif}(0)$  provides the set of initial states whose relaxation is 
exponentially faster than the  set of generic states. Among these initial states, we look for stationary states that can be obtained by applying suitable driving strengths $\xi^2_{0x}$ and $\xi^2_{0y}$. One such example is illustrated in Fig.~(\ref{fig15}). Here, $P$ which is hotter cools exponentially faster.

In these calculations, the spatial degrees of freedom have been ignored. To show that the results continue to hold even when spatial correlations may be present, we compare the analytical results with event-driven molecular dynamics (MD) simulations. In the MD simulations, we have analyzed the systems with number density, $n=$0.02 and $r=0.65$. During a  collision, the velocities are updated according to Eq.~(\ref{eq:collision}). For driving, after a certain time step $dt$, a particle is chosen at random at rate $\lambda_d$ and the velocity of the particle is updated according to: $v'_i=-v_i+\sqrt{\xi^2_{0i}/\lambda_d}\phi_i$, where $i=x, y$ and $\phi$ is drawn from a normal distribution. We prepare two systems $P$ and $Q$ in their stationary state initial conditions having different initial temperatures and then quenched to the same lower temperature. 

\begin{figure}
\centering
\includegraphics[width=\columnwidth]{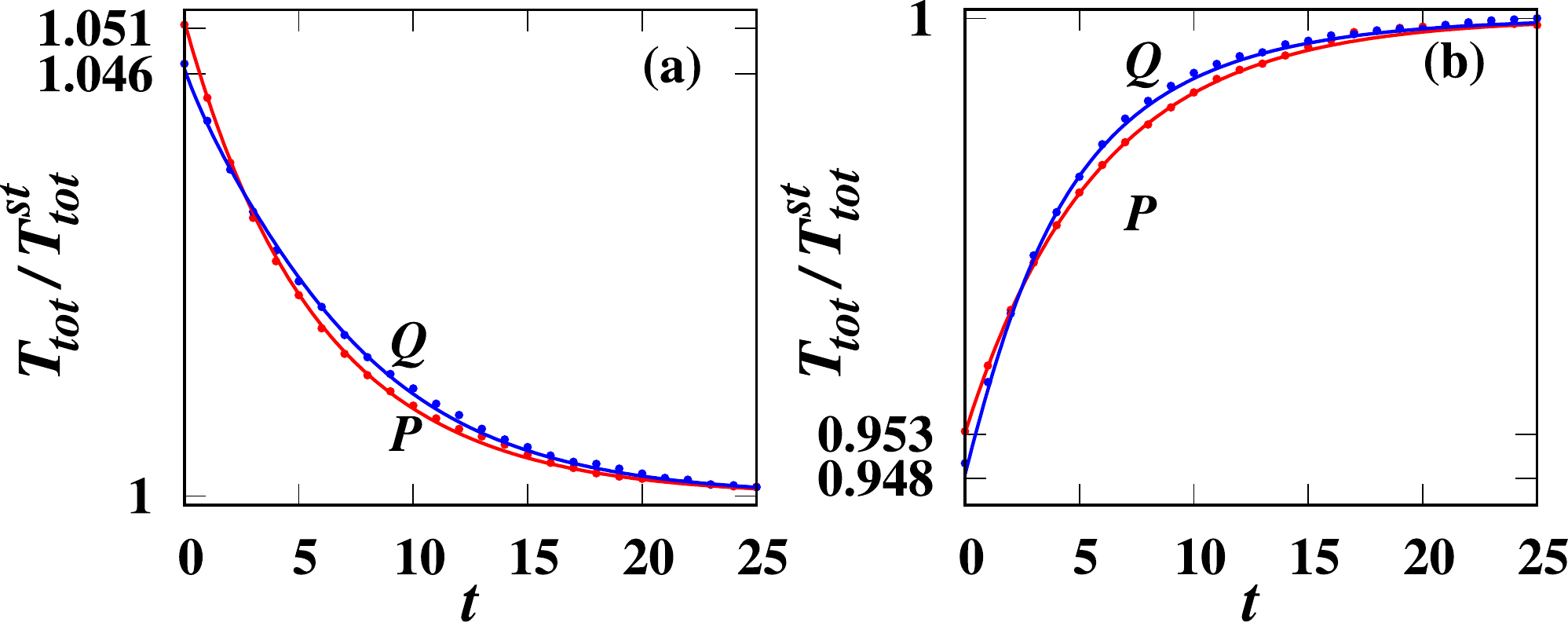}
\caption{Time evolution of $T_{tot}(t)$ with $t$ for two systems $P$ and $Q$ with  initial conditions $T^P_{tot}(0)$=61.0, $T^P_{dif}(0)$=17.08, $T^Q_{tot}(0)$=60.7 and $T^Q_{dif}(0)$=$-10.0$.  (a) The trajectories cross showing the Mpemba effect when $P$ and $Q$ are quenched to the final stationary state values of $T^{st}_{tot}$=58.0 and $T^{st}_{dif}$=16.23. (b)  The same initial conditions show the inverse Mpemba effect when quenched to the final stationary state values of $T^{st}_{tot}$=64.0 and $T^{st}_{dif}$=17.92. The solid lines represent the exact time evolution of $T_{tot}$ and the points represent the results from simulation. The other parameters used for the systems are $m$=1, $n$=0.02 and $r$=0.65.} \label{fig simulation}
\end{figure}

Figure~\ref{fig simulation} shows the time evolution of the total temperature, $T_{tot}$ with time, $t$ when the two systems, $P$ and $Q$ are driven from their different stationary state initial conditions to a same final state. The solid lines represent the theoretical predictions as obtained by solving the full non-linear Eq.~(\ref{time ev}) by assuming Gaussian distribution for the velocity distribution function whereas the points denote the results obtained from the MD simulations. Clearly, there is a good agreement between the theoretical predictions and the results from the MD simulations for both the Mpemba effect [see  Fig.~\ref{fig simulation}(a)] and its inverse [see  Fig.~\ref{fig simulation}(b)].  
For much larger initial temperatures (twice as large), the comparison between simulation and the numerical solution of Eq.~(\ref{time ev}) is shown  in Fig.~(S1) of SM~\cite{supplemental}. While there is a small quantitative difference, the qualitative features of the Mpemba effect persist.

To summarize, we showed the existence of the Mpemba effect, the inverse Mpemba effect and the strong Mpemba effect in an anisotropically driven granular gas.  The key feature is that that the initial states are stationary states unlike earlier analysis of the Mpemba effect in granular systems which required initial states that were not stationary. Our analysis also shows that anisotropy in the velocity distribution of particles is a key ingredient for the existence of such anomalous behaviour, and the Mpemba effect vanishes when the collisions become elastic. Though the exact results were based on a linearized theory for a spatially homogeneous system, MD simulations of hard discs in two dimensions show that the results are true for a spatially extended system also. Achieving anisotropic driving in experiments is not difficult as the amplitude and frequency of shaking can be chosen to be different in different directions, and should therefore allow for an experimental realization, which has been hitherto lacking, of the Mpemba effect in granular systems.


%

\end{document}


\title{Supplemental Material for \\ ``Mpemba effect in an anisotropically driven granular gas''}

\author{Apurba Biswas}
\email{apurbab@imsc.res.in}
\affiliation{The Institute of Mathematical Sciences, C.I.T. Campus, Taramani, Chennai 600113, India}
\affiliation{Homi Bhabha National Institute, Training School Complex, Anushakti Nagar, Mumbai 400094, India}
\author{V . V. Prasad}
\email{prasad.vv@cusat.ac.in}
\affiliation{Government Arts and Science College, Nadapuram, Kozhikode 673506, India}
\affiliation{Department of Physics, Cochin University of Science and Technology,  Kochi 682022, India}
\author{R. Rajesh} 
\email{rrajesh@imsc.res.in}
\affiliation{The Institute of Mathematical Sciences, C.I.T. Campus, Taramani, Chennai 600113, India}
\affiliation{Homi Bhabha National Institute, Training School Complex, Anushakti Nagar, Mumbai 400094, India}

\maketitle
\appendix

In this Supplemental Material (SM), we derive the equations for the time evolution of granular temperatures for an anisotropically driven granular gas. Then we linearize the non-linear time evolution equations by considering initial states that are close to the final stationary state allowing for the analytical calculations to be more tractable. In the last section of SM, we compare the theoretical predictions with event-driven molecular dynamics (MD) simulations for large initial temperatures (compared to final stationary state) that shows the Mpemba effect.

\section{ Derivation for time evolution of temperatures \label{Sup_appendix time ev}}

In this section, we derive the equations for the time evolution of granular temperatures. We define the velocity distribution function $f(\boldsymbol{v},t)$ as the number density of particles having velocity $\boldsymbol{v}$ at time $t$ and it obeys Enskog-Boltzmann equation
\begin{align}
\frac{\partial}{\partial t}f(\boldsymbol{{v}},t)=\chi I(f,f) + \Big(\frac{\xi^2_{0x}}{2} \frac{\partial^2}{\partial v^2_x} + \frac{\xi^2_{0y}}{2} \frac{\partial^2}{\partial v^2_y}\Big) f(\boldsymbol{{v}},t), \label{S_boltzmann eq}
\end{align}
where  $\chi$~\cite{brilliantov2010kinetic} is the pair correlation function, $\xi^2_{0x}$ and $\xi^2_{0y}$ are the variance or strengths of the white noise along the $x$- and $y$-directions respectively and the collision integral  $I(f,f)$ is given by
\bea
I(f,f)=\sigma \int d \boldsymbol{{v}}_2 \int d \boldsymbol{{e}} \Theta(-\boldsymbol{{v}}_{12}.\boldsymbol{{e}}) |\boldsymbol{{v}}_{12}.\boldsymbol{{e}}| \Big[ \frac{1}{r^2} f(\boldsymbol{v}^{''}_1,t) f(\boldsymbol{v}^{''}_2,t) - f(\boldsymbol{{v}}_1,t) f(\boldsymbol{{v}}_2,t)  \Big]. \label{collision integral}
\eea
Here, $(\boldsymbol{v}^{''}_1,\boldsymbol{v}^{''}_2)$ are the pre-collision velocities that give $(\boldsymbol{v}_1, \boldsymbol{v}_2)$ upon collision. In the Gaussian approximation for $f(\boldsymbol{v},t)$(see main text), the time evolutions of $T_{x}(t)$ and $T_{y}(t)$ which are defined as $T_i(t)=(2/n)\int d \boldsymbol{v} \frac{1}{2} m v^2_i  f(\boldsymbol{{v}},t)$, $i\in(x,y)$, can be computed using Eq.~(\ref{S_boltzmann eq}). The time evolution equations are given by
{\footnotesize
\bea
\frac{\partial T_x}{\partial t}=m \xi^2_{0x} + \frac{4 n \chi \sigma (1+r)  \sqrt{T_y}}{15  \sqrt{m} \sqrt{\pi} (T_x - T_y)} \times \Big[ & \Big( 4(2r-3) T^2_x + (7-3r) T_x T_y -2(1+r) T^2_y \Big) {\textit{E}}\Big(1-\frac{T_x}{T_y}\Big) \nonumber \\
& +  T_x \big( 2T_x(3-2r) + T_y(1+r)  \big) {\textit{K}}\Big(1-\frac{T_x}{T_y}\Big)   \Big], \label{time ev Tx 2}
\eea}
and
{\footnotesize
\bea
\frac{\partial T_y}{\partial t}=m \xi^2_{0y} + \frac{4 n \chi \sigma (1+r)  \sqrt{T_y}}{15  \sqrt{m} \sqrt{\pi} (T_x - T_y)}  
\times \Big[ & 2(1+r) (T^2_x - T^2_y) {\textit{E}}\Big(1-\frac{T_x}{T_y}\Big) + (3r-7) \sqrt{T_x T_y} (T_x -2T_y) {\textit{E}}\Big(1-\frac{T_y}{T_x}\Big) \nonumber \\
& -(1+r) T_x (T_x -T_y) {\textit{K}}\Big(1-\frac{T_x}{T_y}\Big) + (3r-7) \sqrt{T_x} T^{3/2}_y {\textit{K}}\Big(1-\frac{T_y}{T_x}\Big)   \Big], \label{time ev Ty 2}
\eea}
where ${\textit{K}}(x)$ and ${\textit{E}}(x)$ are elliptic integrals of first and second kind respectively. We define $T_{tot}(t)=T_x(t) + T_y(t)$ and $T_{dif}(t)=T_x(t) - T_y(t)$. The time evolutions for $T_{tot}(t)$ and $T_{dif}(t)$ can be obtained using Eqs.~(\ref{time ev Tx 2}) and (\ref{time ev Ty 2}) as 
\bea
\begin{split}
\frac{\partial }{\partial t} T_{tot}(t)=\mathcal{F}(T_{tot},T_{dif}),\\
\frac{\partial }{\partial t}T_{dif}(t)=\mathcal{G}(T_{tot},T_{dif}),
\end{split} \label{S_time ev}
\eea

where

{\footnotesize
\bea
\mathcal{F}(T_{tot},T_{dif}) = {m (\xi^2_{0x} + \xi^2_{0y})} + \frac{n \chi \sigma (1+r) \sqrt{T_{tot} - T_{dif}} }{15  \sqrt{2 \pi m} T_{dif}}  \times \Big[& \Big((3r-7)T^2_{tot} + 4(7r-3) T_{tot} T_{dif} + 3(3r-7)T^2_{dif} \Big) {\textit{E}}\Big(\frac{-2T_{dif}}{T_{dif} - T_{dif}}\Big) \nonumber \\
&-(3r-7)(T_{tot} - 3T_{dif}) \sqrt{T^2_{tot} - T^2_{dif}} \textit{{E}}\Big(\frac{2T_{dif}}{T_{tot} + T_{dif}}\Big) \nonumber \\
& - (T_{tot} + T_{dif}) \Big((3r-7) T_{tot} + (7r-3) T_{dif} \Big)  \textit{{K}}\Big(\frac{-2T_{dif}}{T_{tot} - T_{dif}}\Big) \nonumber \\ 
& + (3r-7) \sqrt{T_{tot} + T_{dif}} (T_{tot} - T_{dif})^{3/2}  \textit{{K}}\Big(\frac{2T_{dif}}{T_{tot} + T_{dif}}\Big)  \Big], \label{time ev Tt}
\eea}
and
{\footnotesize
\bea
\mathcal{G}(T_{tot},T_{dif}) ={m(\xi^2_{0x} - \xi^2_{0y})} + \frac{ n \chi \sigma (1+r) (3r-7) \sqrt{T_{tot} - T_{dif}} }{15  \sqrt{2 \pi m} T_{dif}}  \times &\Big[(T_{tot} + 3T_{dif})(T_{tot} + T_{dif}) \textit{{E}}\Big(\frac{-2T_{dif}}{T_{tot} - T_{dif}}\Big) \nonumber \\
&+(T_{tot} - 3T_{dif}) \sqrt{T^2_{tot} - T^2_{dif}} \textit{{E}}\Big(\frac{2T_{dif}}{T_{tot} + T_{dif}}\Big) \nonumber \\
& - (T_{tot} + T_{dif})^2   \textit{{K}}\Big(\frac{-2T_{dif}}{T_{tot} - T_{dif}}\Big) \nonumber \\ 
& - \sqrt{T_{tot} + T_{dif}} (T_{tot} - T_{dif})^{3/2}  \textit{{K}}\Big(\frac{2T_{dif}}{T_{tot} + T_{dif}}\Big)  \Big]. \label{time ev Td}
\eea}

\section{ Linearized time evolution equations \label{appendix linearized theory}}

In this section, we show that the non-linear time evolution equations in Eq.~(\ref{S_time ev}) can be linearized by considering initial states that are close to final stationary state denoted by $T^{st}_{tot}$ and $T^{st}_{dif}$. Given that, $\delta T_{tot}(t)=T_{tot}(t)- T^{st}_{tot}$ and $\delta T_{dif}(t)=T_{dif}(t) - T^{st}_{dif}$, the linearized time evolution of $\delta T_{tot}$ and $\delta T_{dif}$ can be written in a matrix form {\footnotesize $\frac{d}{dt}\begin{pmatrix}
\delta T_{tot}(t) \\
\delta T_{dif}(t)
\end{pmatrix}=-~\boldsymbol{R}\begin{pmatrix}
\delta T_{tot}(t) \\
\delta T_{dif}(t)
\end{pmatrix}$}, where $\boldsymbol{R}$ is a $2\times2$ matrix with components $R_{11}$, $R_{12}$, $R_{21}$ and $R_{22}$ given by
{\footnotesize
\bea
R_{11}=&-\frac{n \chi \sigma (1+r) }{30  \sqrt{2 \pi m} T^{st}_{dif}\sqrt{T^{st}_{tot} - T^{st}_{dif}}} \times
\Big[((3 - 47 r) (T^{st}_{dif})^2 + 8 (9r -1) T^{st}_{dif} T^{st}_{tot} + 5( 3 r-7) (T^{st}_{tot})^2)  \textit{{E}}\Big(\frac{-2T^{st}_{dif}}{T^{st}_{tot} - T^{st}_{dif}}\Big) \nonumber \\ 
&+\frac{7-3r}{\sqrt{(T^{st}_{tot})^2 - (T^{st}_{dif})}}\Big(5 (T^{st}_{dif})^2 + 3 (T^{st}_{dif})^2 T^{st}_{tot} -13 T^{st}_{dif}(T^{st}_{tot})^2 + 5 (T^{st}_{tot})^2 \Big) \textit{{E}}\Big(\frac{2T^{st}_{dif}}{T^{st}_{tot} + T^{st}_{dif}}\Big)  \nonumber \\
&+ \frac{(7-3r) \sqrt{T^{st}_{tot} - T^{st}_{dif}}}{\sqrt{T^{st}_{tot} + T^{st}_{dif}}} \Big(T^{st}_{dif} (-4\sqrt{T_{tot} + T_{dif}} +1) - T^{st}_{tot}(2\sqrt{T_{tot} + T_{dif}} +3) \Big)  \textit{{K}}\Big(\frac{2T^{st}_{dif}}{T^{st}_{tot} + T^{st}_{dif}}\Big) \nonumber \\
& +\Big((13r-17)(T^{st}_{dif})^2 + 2(1-9r)T^{st}_{dif} T^{st}_{tot} + 5(7-3r)(T^{st}_{tot})^2 \Big)  \textit{{K}}\Big(\frac{-2T^{st}_{dif}}{T^{st}_{tot} - T^{st}_{dif}}\Big)  \nonumber \\
& +\frac{\pi T^{st}_{dif} }{2 (T^{st}_{tot} - T^{st}_{dif})} \Big(3(3r-7)(T^{st}_{dif})^2 + 4(7r-3) T^{st}_{dif} T^{st}_{tot} + (3r-7) (T^{st}_{tot})^2 \Big)  {}_2F_1\Big(\frac{1}{2},\frac{3}{2};2; \frac{-2T^{st}_{dif}}{T^{st}_{tot} - T^{st}_{dif}} \Big) \nonumber \\
& -\frac{\pi (3r-7) T^{st}_{dif} \sqrt{T^{st}_{tot} - T^{st}_{dif}} }{2 (T^{st}_{tot} + T^{st}_{dif})^{3/2}} \Big(3(T^{st}_{dif})^2 - 4 T^{st}_{dif} T^{st}_{tot} +  (T^{st}_{tot})^2 \Big) {}_2F_1\Big(\frac{1}{2},\frac{3}{2};2; \frac{2T^{st}_{dif}}{T^{st}_{tot} + T^{st}_{dif}} \Big) \nonumber \\
& - \frac{\pi T^{st}_{dif} (T^{st}_{tot} + T^{st}_{dif}) }{2 (T^{st}_{tot} - T^{st}_{dif})} \Big((7r-3) T^{st}_{dif} + (3r-7) T^{st}_{tot} \Big) {}_2F_1\Big(\frac{3}{2},\frac{3}{2};2; \frac{-2T^{st}_{dif}}{T^{st}_{tot} - T^{st}_{dif}} \Big)  \nonumber \\
&-\frac{\pi (3r-7) T^{st}_{dif} (T^{st}_{tot} - T^{st}_{dif})^{5/2} }{2 (T^{st}_{tot} + T^{st}_{dif})^{3/2}} {}_2F_1\Big(\frac{3}{2},\frac{3}{2};2; \frac{2T^{st}_{dif}}{T^{st}_{tot} + T^{st}_{dif}} \Big)
\Big], \label{R11} \nonumber
\eea}

{\footnotesize
\bea
R_{12}=&-\frac{n \chi \sigma (1+r) }{30  \sqrt{2 \pi m} (T^{st}_{dif})^2\sqrt{T^{st}_{tot} - T^{st}_{dif}}} \times
\Big[\Big(9(7 - 3 r) (T^{st}_{dif})^3 -10(r+3) (T^{st}_{dif})^2 T^{st}_{tot}-(7-3r)T^{st}_{dif}(T^{st}_{tot})^2 + 2( 7-3 r) (T^{st}_{tot})^3\Big)  \textit{{E}}\Big(\frac{-2T^{st}_{dif}}{T^{st}_{tot} - T^{st}_{dif}}\Big) \nonumber \\ 
&-(7-3r)\frac{(T^{st}_{tot} - T^{st}_{dif})^{3/2}}{\sqrt{T_{tot} + T_{dif}}}\Big(15 (T^{st}_{dif})^2 + 9 T^{st}_{dif} T^{st}_{tot} - 2 (T^{st}_{tot})^2 \Big) \textit{{E}}\Big(\frac{2T^{st}_{dif}}{T^{st}_{tot} + T^{st}_{dif}}\Big)  \nonumber \\
&- \frac{(7-3r) \sqrt{T_{tot} - T_{dif}}}{\sqrt{T_{tot} + T_{dif}}} \Big(5(T^{st}_{dif})^3 - 4 (T^{st}_{dif})^2 T^{st}_{tot}- 3 (T^{st}_{tot})^2 T^{st}_{dif} + 2 (T^{st}_{tot})^3 \Big)  \textit{{K}}\Big(\frac{2T^{st}_{dif}}{T^{st}_{tot} + T^{st}_{dif}}\Big) \nonumber \\
& +\Big(-(3-7r)(T^{st}_{dif})^3 + 4(1+r) (T^{st}_{dif})^2 T^{st}_{tot}+ 5(1-5r) (T^{st}_{tot})^2 T^{st}_{dif} + 2(7-3r)(T^{st}_{tot})^3 \Big)  \textit{{K}}\Big(\frac{-2T^{st}_{dif}}{T^{st}_{tot} - T^{st}_{dif}}\Big)  \nonumber \\
& -\frac{\pi T^{st}_{dif} T^{st}_{tot} }{2 (T^{st}_{tot} - T^{st}_{dif})} \Big(3(7-3r)(T^{st}_{dif})^2 + 4(3-7r) T^{st}_{dif} T^{st}_{tot} + (7-3r) (T^{st}_{tot})^2 \Big)  {}_2F_1\Big(\frac{1}{2},\frac{3}{2};2; \frac{-2T^{st}_{dif}}{T^{st}_{tot} - T^{st}_{dif}} \Big) \nonumber \\
& -\frac{\pi (7-3r) T^{st}_{dif} T^{st}_{tot} \sqrt{T^{st}_{tot} - T^{st}_{dif}} }{2 (T^{st}_{tot} + T^{st}_{dif})^{3/2}} \Big(3(T^{st}_{dif})^2 - 4 T^{st}_{dif} T^{st}_{tot} +  (T^{st}_{tot})^2 \Big) {}_2F_1\Big(\frac{1}{2},\frac{3}{2};2; \frac{2T^{st}_{dif}}{T^{st}_{tot} + T^{st}_{dif}} \Big) \nonumber \\
& - \frac{\pi T^{st}_{dif} T^{st}_{tot} (T^{st}_{tot} + T^{st}_{dif}) }{2 (T^{st}_{tot} - T^{st}_{dif})} \Big((3-7r) T^{st}_{dif} + (7-3r) T^{st}_{tot} \Big) {}_2F_1\Big(\frac{3}{2},\frac{3}{2};2; \frac{-2T^{st}_{dif}}{T^{st}_{tot} - T^{st}_{dif}} \Big)  \nonumber \\
&-\frac{\pi (7-3r) T^{st}_{dif} T^{st}_{tot} (T^{st}_{tot} - T^{st}_{dif})^{5/2} }{2 (T^{st}_{tot} + T^{st}_{dif})^{3/2}} {}_2F_1\Big(\frac{3}{2},\frac{3}{2};2; \frac{2T^{st}_{dif}}{T^{st}_{tot} + T^{st}_{dif}} \Big)
\Big], \label{R12} \nonumber
\eea}

{\footnotesize
\bea
R_{21}=&\frac{n \chi \sigma (1+r) (7-3r)}{30  \sqrt{2 \pi m} T^{st}_{dif}\sqrt{T^{st}_{tot} - T^{st}_{dif}}} \times 
\Big[\Big(-5 (T^{st}_{dif})^2 + 8  T^{st}_{dif} T^{st}_{tot} + 5 (T^{st}_{tot})^2 \Big)  \textit{{E}}\Big(\frac{-2T^{st}_{dif}}{T^{st}_{tot} - T^{st}_{dif}}\Big) \nonumber \\ 
&+\Big(\frac{(T^{st}_{tot} + T^{st}_{dif})^2}{2}+ (3T^{st}_{tot} - T^{st}_{dif})(T^{st}_{tot} -3 T^{st}_{dif}) \frac{\sqrt{T^{st}_{tot} - T^{st}_{dif}}}{\sqrt{T^{st}_{tot} + T^{st}_{dif}}}\Big) \textit{{E}}\Big(\frac{2T^{st}_{dif}}{T^{st}_{tot} + T^{st}_{dif}}\Big)  \nonumber \\
&-\Big(\frac{4}{\sqrt{(T^{st}_{tot})^2 - (T^{st}_{dif})}} \big(4 (T^{st}_{dif})^2 -  T^{st}_{dif} T^{st}_{tot} - (T^{st}_{tot})^2 \big) \Big) \textit{{E}}\Big(\frac{2T^{st}_{dif}}{T^{st}_{tot} + T^{st}_{dif}}\Big)  \nonumber \\
&- \frac{ (T^{st}_{tot} - T^{st}_{dif})^{3/2}}{\sqrt{T^{st}_{tot} + T^{st}_{dif}}} \Big(3 T^{st}_{dif} + 5 T^{st}_{tot} \Big)  \textit{{K}}\Big(\frac{2T^{st}_{dif}}{T^{st}_{tot} + T^{st}_{dif}}\Big) -\Big((5T^{st}_{tot}-3 T^{st}_{dif}) (T^{st}_{tot}+ T^{st}_{dif})\Big)  \textit{{K}}\Big(\frac{-2T^{st}_{dif}}{T^{st}_{tot} - T^{st}_{dif}}\Big)  \nonumber \\
& -\frac{\pi T^{st}_{dif}  (T^{st}_{tot} +3 T^{st}_{dif})  (T^{st}_{tot} + T^{st}_{dif})} {2 (T^{st}_{tot} - T^{st}_{dif})}   {}_2F_1\Big(\frac{1}{2},\frac{3}{2};2; \frac{-2T^{st}_{dif}}{T^{st}_{tot} - T^{st}_{dif}} \Big)  +\frac{\pi T^{st}_{dif}  (T^{st}_{tot} -3 T^{st}_{dif})(T^{st}_{tot} - T^{st}_{dif})^{3/2} }{2 (T^{st}_{tot} + T^{st}_{dif})^{3/2}}  {}_2F_1\Big(\frac{1}{2},\frac{3}{2};2; \frac{2T^{st}_{dif}}{T^{st}_{tot} + T^{st}_{dif}} \Big) \nonumber \\
& -\frac{\pi T^{st}_{dif} (T^{st}_{tot} + T^{st}_{dif})^2} {2 (T^{st}_{tot} - T^{st}_{dif})} {}_2F_1\Big(\frac{3}{2},\frac{3}{2};2; \frac{-2T^{st}_{dif}}{T^{st}_{tot} - T^{st}_{dif}} \Big)  +\frac{\pi T^{st}_{dif} (T^{st}_{tot} - T^{st}_{dif})^{5/2}} {2 (T^{st}_{tot} + T^{st}_{dif})^{3/2}} {}_2F_1\Big(\frac{3}{2},\frac{3}{2};2; \frac{2T^{st}_{dif}}{T^{st}_{tot} + T^{st}_{dif}} \Big)
\Big], \label{R21} \nonumber
\eea}

and 
{\footnotesize
\bea
R_{22}=&\frac{n \chi \sigma (1+r)(7-3r) }{30  \sqrt{2 \pi m} (T^{st}_{dif})^2\sqrt{T^{st}_{tot} - T^{st}_{dif}}} \times
\Big[\Big(-9 (T^{st}_{dif})^3 + 2 (T^{st}_{dif})^2 T^{st}_{tot}+T^{st}_{dif}(T^{st}_{tot})^2 - 2 (T^{st}_{tot})^3\Big)  \textit{{E}}\Big(\frac{-2T^{st}_{dif}}{T^{st}_{tot} - T^{st}_{dif}}\Big) \nonumber \\ 
&+\frac{\sqrt{T^{st}_{tot} - T^{st}_{dif}}}{\sqrt{T_{tot} + T_{dif}}}\Big(9 (T^{st}_{dif})^3 + 2 (T^{st}_{dif})^2 T^{st}_{tot}-T^{st}_{dif}(T^{st}_{tot})^2 - 2 (T^{st}_{tot})^3 \Big) \textit{{E}}\Big(\frac{2T^{st}_{dif}}{T^{st}_{tot} + T^{st}_{dif}}\Big)  \nonumber \\
&+ \frac{ (T_{tot} - T_{dif})^{3/2}}{\sqrt{T_{tot} + T_{dif}}} \Big(3 (T^{st}_{dif})^2 + 3 T^{st}_{dif} T^{st}_{tot}  + 2 (T^{st}_{tot})^2 \Big)  \textit{{K}}\Big(\frac{2T^{st}_{dif}}{T^{st}_{tot} + T^{st}_{dif}}\Big) \nonumber \\
& +\Big(3 (T^{st}_{dif})^2 - 3 T^{st}_{dif} T^{st}_{tot}  + 2 (T^{st}_{tot})^2 \Big)  \textit{{K}}\Big(\frac{-2T^{st}_{dif}}{T^{st}_{tot} - T^{st}_{dif}}\Big)  \nonumber \\
& +\frac{\pi T^{st}_{dif} T^{st}_{tot} (T^{st}_{tot} + T^{st}_{dif}) (T^{st}_{tot} + 3T^{st}_{dif}) }{2 (T^{st}_{tot} - T^{st}_{dif})}   {}_2F_1\Big(\frac{1}{2},\frac{3}{2};2; \frac{-2T^{st}_{dif}}{T^{st}_{tot} - T^{st}_{dif}} \Big) \nonumber \\
& -\frac{\pi T^{st}_{dif} T^{st}_{tot} (T^{st}_{tot} - 3 T^{st}_{dif}) (T^{st}_{tot} - T^{st}_{dif})^{3/2}} {2 (T^{st}_{tot} + T^{st}_{dif})^{3/2}} {}_2F_1\Big(\frac{1}{2},\frac{3}{2};2; \frac{2T^{st}_{dif}}{T^{st}_{tot} + T^{st}_{dif}} \Big) \nonumber \\
& +\frac{\pi T^{st}_{dif} T^{st}_{tot} (T^{st}_{tot} + T^{st}_{dif})^2 }{2 (T^{st}_{tot} - T^{st}_{dif})} {}_2F_1\Big(\frac{3}{2},\frac{3}{2};2; \frac{-2T^{st}_{dif}}{T^{st}_{tot} - T^{st}_{dif}} \Big)  \nonumber \\
&-\frac{\pi T^{st}_{dif} T^{st}_{tot} (T^{st}_{tot} - T^{st}_{dif})^{5/2}} {2 (T^{st}_{tot} + T^{st}_{dif})^{3/2}} {}_2F_1\Big(\frac{3}{2},\frac{3}{2};2; \frac{2T^{st}_{dif}}{T^{st}_{tot} + T^{st}_{dif}} \Big)
\Big]. \label{R22}
\eea}
The solutions for the time evolution of $\delta T_{tot}$ and $\delta T_{dif}$ are:
\bea
\begin{split}
\delta T_{tot}(t)=K_+ e^{-\lambda_+ t} + K_- e^{-\lambda_- t}, \\
\delta T_{dif}(t)=L_+ e^{-\lambda_+ t} + L_- e^{-\lambda_- t}, \label{S_time ev delta Tt}
\end{split}
\eea 
where the coefficients $K_+, K_-, L_+$ and $L_-$ are given by
\bea
\begin{split}
&K_+=\frac{1}{\gamma}\Big[R_{12} \delta T_{dif}(0)  - (\lambda_- - R_{11}) \delta T_{tot}(0) \Big], \\
&K_-=\frac{1}{\gamma}\Big[-R_{12} \delta T_{dif}(0)  + (\lambda_+ - R_{11}) \delta T_{tot}(0) \Big], \\
&L_+=\frac{1}{\gamma}\Big[(\lambda_+ - R_{11}) \delta T_{dif}(0)  - \frac{(\lambda_+ - R_{11})(\lambda_- - R_{11})}{R_{12}} \delta T_{tot}(0) \Big], \\
&L_-=\frac{1}{\gamma}\Big[-(\lambda_- - R_{11}) \delta T_{dif}(0)  + \frac{(\lambda_+ - R_{11})(\lambda_- - R_{11})}{R_{12}} \delta T_{tot}(0) \Big].
\end{split} \label{coefficients}
\eea

\section{ Mpemba effect for large initial temperatures \label{appendix larger temperature}}

\begin{figure}
\centering
\includegraphics[width=8.6 cm]{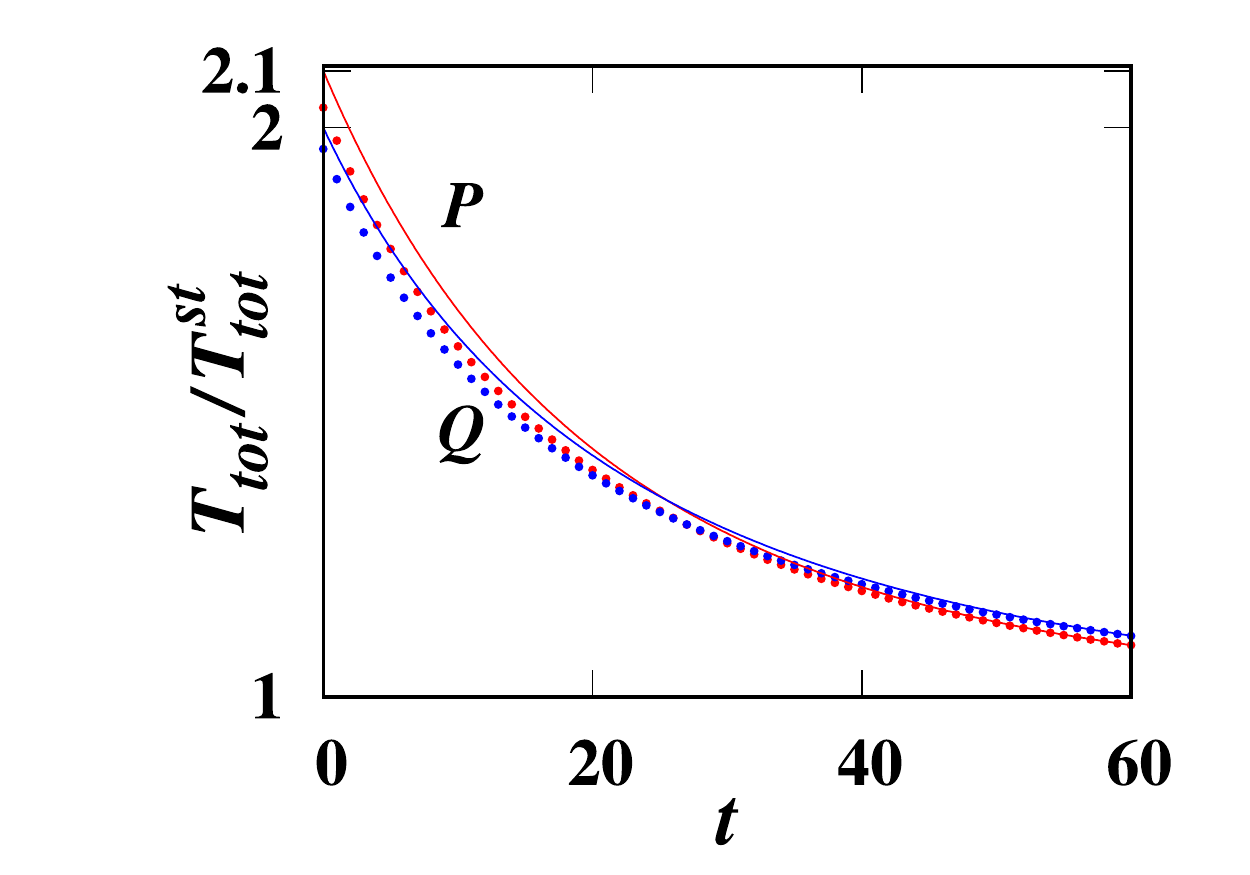}
\caption{Time evolution of $T_{tot}(t)$ with $t$ for two identical systems $P$ and $Q$ with  initial conditions in (a) $T^P_{tot}(0)$=2.1, $T^P_{dif}(0)$=1.26, $T^Q_{tot}(0)$=2.0 and $T^Q_{dif}(0)$=$-1.2$ show the Mpemba effect when quenched to the final stationary state values of $T^{st}_{tot}$=1.0 and $T^{st}_{dif}$=0.6. The other parameters used for the system are $r=0.05$, $m=1$ and $n=0.02$. The solid lines represent the exact time evolution of $T_{tot}$ and the points represent the results from simulation. }\label{fig mpemba_small_r}
\end{figure}

In this section, we illustrate that even large initial temperatures (twice as large) compared to the final steady state, can also lead to the Mpemba effect. For this we consider very inelastic system with coefficient of restitution $r=0.05$ for a number density of particles, $n=0.02$. Figure~(\ref{fig mpemba_small_r}) compares the theoretical predictions (solid lines) as obtained by numerically solving Eqs.~(\ref{S_time ev}) by assuming Gaussian distribution for the velocity distribution function with the results obtained from the detailed MD simulations (points). It  shows that although there is a quantitative difference, the qualitative results of the Mpemba effect remain the same. 

